# Методы исследования и свойства МДП-структур на основе кремния

Я. Г. Федоренко

Stephenson Institute for Renewable Energy and Department of Physics, School of Physical Sciences, Chadwick Building, University of Liverpool, Liverpool, UK

**Аннотация**

В работе проведен обзор методов, применяющихся в изучении электронных и структурных свойств границы раздела полупроводник-диэлектрик. Рассмотрены методы определения параметров поверхностных состояний полупроводника и методы исследования захваченного в окисле заряда. Проведена классификация основных структурных дефектов системы Si−SiO$_2$.

**Silicon-Oxide Interfaces: Structure and Electronic Properties**

Y. G. Fedorenko

Stephenson Institute for Renewable Energy and Department of Physics, School of Physical Sciences, Chadwick Building, University of Liverpool, Liverpool, UK

Abstract

The paper reviews the methods used to study the electronic and structural properties of the semiconductor-dielectric interfaces. Methodological approaches to study the interface states and charge trapping in the oxide are considered. An overview of archetypical structural defects of the Si–SiO$_2$ interface is given.

## 1. Введение

Граница раздела полупроводник-диэлектрик является гетерогенной системой, электронные и структурные свойства которой существенным образом зависят от технологии ее формирования. Структурные несовершенства и модификация состава тонких диэлектрических пленок на полупроводниках определяют характеристики устройств на основе структур металл-диэлектрик-полупроводник (МДП). Данная статья рассматривает методы исследования, использующиеся при изучении электронных свойств границы раздела диэлектрик-полупроводник, и их зависимость от наличия точечных дефектов в окисле и в области интерфейса.

## 2. Структура дефектов границы раздела Si/SiO$_2$

Поскольку система Si/SiO$_2$ характеризуется наличием парамагнитных дефектов, обусловленных появлением оборванных связей кремния и кислорода, исследование микроструктуры этих дефектов проводилось методами спектроскопии электронного парамагнитного резонанса [1]. Физические факторы, определяющие спектр поглощения микроволнового излучения, описываются спиновым гамильтонианом:

$$H = \beta \cdot B \cdot g \cdot S + I \cdot A \cdot S, \qquad (1)$$

где первая часть выражения соответствует взаимодействию Зеемана, а вторая – сверхтонкому взаимодействию между неспаренным спином электрона и спином ядра. Сверхтонкое взаимодействие обычно содержит изотропную компоненту, являющуюся мерой концентрации волновой функции ядра, и анизотропную, возникающую из-за диполь-дипольных взаимодействий между электронным и ядерным спином. Основными параметрами, измеряемыми экспериментально, являются проекции **g**-фактора (фактора спектроскопического расщепления) и проекции тензора сверхтонкого взаимодействия **A** в трехмерном пространстве координат. Анализ сверхтонкого взаимодействия позволяет идентифицировать атомы и их расположение в веществе. Ориентация и симметрия тензоров **g** и **A** отражает влияние волновой функции, определяемой локальным окружением дефекта, и дает информацию о пространственной ориентации дефекта в кристалле. В некристаллическом твердом теле анизотропные спектральные свойства пространственно усреднены, что приводит к размытию спектра. Для эксперимента характерно влияние нескольких факторов на форму сигнала и ширину спектральных линий. В последнем случае необходимо принимать во внимание взаимодействие между спином и его окружением (спин-дипольные взаимодействие) и влияние механических деформаций. Ширина резонансной линии определяется двумя факторами – спин-решеточной релаксацией и спин-спиновой релаксацией. Первая компонента возникает из-за релаксации парамагнитной системы в магнитном поле к равновесному состоянию и зависит от температуры. Вторая компонента определяется концентрацией спинов и характеризует осциллирующую составляющую намагниченности в плоскости, перпендикулярной направлению магнитного поля. Ширина резонансной линии обратно пропорциональна временам релаксации, характеризующим данные процессы.

Классификация ЭПР-сигнала осуществляется посредством анализа особенностей спектра – позиции резонансной линии, ширины линии, ее формы и амплитуды. При непрерывном воздействии магнитного поля на образец определяют производную спектра поглощения микроволнового излучения. Условие резонанса при поглощении микроволнового излучения в магнитном поле описывается выражением $h\nu = \mathbf{g}\cdot\beta\cdot\mathbf{B}$, где $h$ and $\beta$ есть постоянная Планка и магнетон Бора, соответственно, а $\nu$ – частота высокочастотного сигнала. Внешнее магнитное поле $\mathbf{B}$ измеряется непосредственно на исследуемом образце, для чего применяется протонный зонд, помещенный в полость резонатора, или образец-маркер, значение **g** которого заранее определено. Часто таким образцом служит сильно легированный фосфором кремний [2]. Значение **g** определяется из компоненты Зеемана, входящей в гамильтониан. Форма абсорбционной кривой может быть описана фактором k≡$\frac{I}{A\Delta B_{pp}^2}$, где A – ½ амплитуды сигнала $A_{pp}$, *I* определяется двойным интегрированием производной спектра поглощения микроволнового излучения.

Рассмотрим основные парамагнитные дефекты системы $Si/SiO_2$. Термическое окисление кремния является результатом появления общих валентных электронов у кремния и кислорода и сопровождается генерацией дефектов оборванных связей на ГР $Si/SiO_2$. Часть дефектов, такие как $P_b$ дефекты, парамагнитны. Полагают, что дефекты границы раздела возникают при компенсации механических напряжений из-за несоответствия постоянных решеток. Структура $P_b$ дефектов зависит от кристаллографической ориентации кремния. Для (111)$Si/SiO_2$ интерфейса характерно наличие только одного типа дефекта оборванных связей – $P_b$ центра. Его конфигурация показана на рис. 2. Этот дефект был идентифицирован как *sp3* оборванная связь атома кремния в направлении [111]. Дефект имеет $C_{3v}$ симметрию по отношению к оси [111] [3, 4]. При окислении кремния в интервале температур 300-950°C плотность $P_b$ дефектов не изменяется и составляет $4.9 \cdot 10^{12}$ см$^{-2}$. Для (100)$Si/SiO_2$ системы характерно наличие двух типов парамагнитных дефектов, $P_{b0}$ и $P_{b1}$, Рис. 3. При окислении кремния в интервале температур 800-970°C плотность $P_{b0}$ и $P_{b1}$ дефектов примерно одинакова и равна $10^{12}$ см$^{-2}$. $P_{b0}$ и $P_{b1}$ характеризуются пониженной симметрией по сравнению с $P_b$ ($C_2$, а не $C_{3v}$), хотя симметрия $P_{b0}$ аксиальна к направлению <111>, и значения фактора **g** для дефектов $P_b$ и $P_{b0}$ близки [5]. Модель $P_{b1}$ дефекта может быть представлена неспаренной связью атома кремния, который связан с тремя атомами кремния, но оборванная связь кремния смещена к объему кремния от ГР (100)$Si/SiO_2$. Неспаренный спин $P_{b1}$ дефекта может быть направлен к стороннему атому, что понижает симметрию дефекта.

Исследование дефектов оборванных связей кремния ориентации (100) и взаимосвязь между дефектами оборванных связей кремния и электронными состояниями ГР Si/SiO$_2$ начато в работах [6, 7]. Исследования электрической активности дефектов оборванных связей кремния позволили установить однозначную количественную корреляцию между плотностью $P_b$ и плотностью поверхностных состояний ГР Si/SiO$_2$. Исследования проводились с использованием различных методик для определения плотности поверхностных состояний: вольт-емкостных измерений [8], релаксационной спектроскопии глубоких уровней [9] и по определению порога фотоионизации [10]. На основе определения плотности $P_{b0}$ и $P_{b1}$ дефектов ГР (100)Si/SiO$_2$ методом ЭПР и корреляции с плотностью поверхностных состояний, полученной на основе анализа вольт-фарадных характеристик (ВФХ), было показано, что дефекты $P_{b1}$ типа электрически активных поверхностных состояний в запрещенной зоне кремния не образуют [11]. Дефекты $P_{b0}$ образуют амфотерные поверхностные состояния при энергиях 0.3 eB и 0.8 eB относительно края валентной зоны. В работе [12] была установлена прямая корреляция между плотностью $P_b$ дефектов и концентрацией свободных носителей заряда в канале МДП-транзистора. Следует отметить, что на сегодняшний момент развиты технологии создания приповерхностных слоев кремния, основанные на метастабильном встраивании кислорода в кристаллическую решетку кремния методом формирования кислород-кремниевых сверхрешеток. Такой способ формирования монослоев не приводит к возникновению механических напряжений в структуре сверхрешеток, и дефекты оборванных связей кремния в мультислоях кислород-кремний не наблюдаются, хотя и присутствуют на границе раздела кремний-диэлектрик [13]. Используя технологию кислород-кремниевых чередующихся монослоев, при которой атомы кислорода располагаются между связями Si−Si и не являются атомами замещения, можно добиться увеличения подвижности в канале МДП-транзистора, поскольку кислород приводит к искажению кристаллического потенциала кремния, перераспределению подзонных носителей заряда и уменьшению эффективной массы [14].

Дефекты оборванных связей кремния могут исследоваться при изучении деформаций границы раздела диэлектрик-полупроводник (ГР ДП). Пространственная однородность дефектов оборванных связей кремния определяется температурными условиями получения образца, то есть резкостью границы раздела. Определение значений механических деформаций на основе исследования $P_b$-дефектов границы раздела (111)Si/SiO$_2$ было проведено на основе зависимости ширины ЭПР линии от угла магнитного поля [15, 16]. Этот метод может быть применен и к границе раздела (100)Si/SiO$_2$, если плотности $P_{b0}$ и $P_{b1}$ дефектов существенно отличаются. Кроме того, механические деформации могут быть определены по частотной зависимости ЭПР-сигнала при фиксированной ориентации магнитного поля [17]. Поскольку дефекты оборванных связей кремния возникают в результате появления внутренних механических напряжений на границе раздела Si/SiO$_2$, исследовалась зависимость между усредненной величиной механического напряжения из-за несоответствия постоянных решеток и температурой окисления кремния в интервале температур от 22°C до 1140°C [18]. Плотность $P_b$-центров практически не изменяется при температуре окисления от 300°C до 900°C. Если процесс окисления проводится при T>900°C, происходит структурная релаксация границы раздела Si/SiO$_2$, и плотность $P_b$-центров уменьшается. При этом различают две стадии процесса релаксации: при температуре окисления близкой к 900°C имеет место релаксация субоксидных связей границы раздела, а при дальнейшем увеличении температуры происходит уменьшение механических напряжений на макромасштабе [19].

Дефекты, присущие SiO$_2$, принято делить на две группы в зависимости от того, на атоме кремния или кислорода локализована оборванная связь. EX центр относится ко второму типу дефектов и является единственным собственным дефектом в термически окисленном кремнии. Этот дефект образуется в верхней части оксида кремния и наблюдается при окислении кремния в интервале температур $T_{ox}$=700-930°C. Появление EX дефекта связано с перенасыщением кислородом при окислении кремния. Максимальная концентрация EX центров наблюдается в окислах толщиной 100-125 Å. С увеличением толщины окисла концентрация EX дефектов уменьшается. На основе анализа сверхтонкого взаимодействия $^{17}$O было показано, что дефект характеризуется локализацией электрона на нескольких атомах кислорода. Предполагалось, что данному типу дефекта присуща общая составляющая – вакансия кремния [20], рис. 4(а). В парамагнитном состоянии EX центр имеет положительный заряд, который нейтрализуется при взаимодействии с водородом. Другие дефекты окиси кремния вносятся на стадии постростовой технологии и/или при воздействии ионизирующих излучений. К таким дефектам относятся дефект немостикового кислорода (non bridging oxygen hole center (NBOHC)), обозначающийся как ≡Si−O• (рис. 4(б)) [21] и пероксид-радикал, Si−O−O• ((рис. 4(в)) [22, 23]. Возникновение пероксид-радикала при воздействии ионизирующего облучения предполагает наличие *напряженных* Si−O−O−Si связей в образцах аморфного кремния до облучения, поскольку *ненапряженные* связи стабильны, что и наблюдается, если межатомные расстояния Si−Si составляют 4-5 ангстрем. В отсутствие напряженных связей Si−Si, пероксид-радикал может возникать, если предположить, что парамагнитная часть дефекта (=Si$^{+}$−O−O•) и положительный заряд ($^{+}$Si≡) связаны с одним и тем же атомом кислорода и образуют конфигурацию =Si$^{+}$−O−Si−O−O•, как показано на рис. 4(г) согласно работе [24]. В данной конфигурации пероксидная часть дефекта может соединяться только с одним атомом кремния, не приводя к увеличению длины Si−Si связи из-за встраивания атома кислорода.

К дефектам оборванной связи, локализованной на кремнии, относится E' центр, наблюдающийся как в кристаллических, так и аморфных образцах $SiO_2$. Обозначение E' говорит о том, что такой дефект содержит один неспаренный электрон. Электрон локализован на *sp3* орбитали кремния, связанного с тремя атомами кислорода, что обозначается как $O_3\equiv Si\bullet$ [25]. Общим источником формирования E' дефектов является напряженная Si–Si связь. Существует несколько типов E' центров. Рассмотрим два типа дефектов оборванной связи кремния в аморфном $SiO_2$ – $E'_\gamma$ и $E'_\sigma$. Предполагалось, что $E'_\gamma$ подобен дефекту $E'_1$, который был обнаружен в кристаллическом полиморфе оксида кремния — α-кварце (при обозначении E´ дефектов в аморфном твердом теле принято использовать греческую букву как подстрочный знак). Согласно модели дефекта, изображенной на рис. 5(а), в парамагнитном состоянии $E'_\gamma$ положительно заряжен. Положительный заряд, захваченный на нейтральной моновакансии кислорода, локализован на атоме кремния. Релаксация положительно заряженного кремния стабилизируется образованием связи Si–O с соседним кислородом, повышая координацию этого атома кислорода до трех. Более поздней моделью E' дефекта является теоретическая модель бескислородного мостикового центра захвата дырок (the bridged hole-trapping oxygen-deficiency center) [26], рис. 5(б). Согласно этой модели, парамагнитный атом кремния имеет общий атом кислорода с атомом кремния, являющимся центром захвата дырок. Однако данная модель не имеет экспериментального подтверждения.

Центр $E'_\delta$ связан с образованием вакансии кислорода и был обнаружен в образцах аморфного $SiO_2$, содержащих примесь хлора и подвергавшихся рентгеновскому облучению [27]. Спиновая плотность $E'_\delta$, определенная по величине сверхтонкого расщепления $Si^{29}$, почти в четыре раза меньше, чем для $E'_\gamma$, для которого примерно 85% спиновой плотности сосредоточено на *sp3* орбитали одного атома кремния. Поэтому авторы предположили, что новый дефект

E'$_\delta$ содержит неспаренный спин, делокализованный на четырех атомах кремния. При исследовании E' дефектов в окисле кремния, изготовленном по технологии кремний-на-изоляторе, механизм образования E'$_\delta$ не связан с наличием Cl или F, а обусловлен образованием кремниевых кластеров, хотя примеси Cl или F, могут приводить к усилению кластерообразования [28]. Модель дефекта, предполагающая делокализацию спина электрона на четырех атома кремния, которые соединены с пятым центральным атомом кремния, показана на рис. 5(в).

Зарядовое состояние E' центров исследовалось в работе [29]. Показано, что корреляция между плотностью E' центров и положительным зарядом в окисле не наблюдается. В парамагнитном состоянии E' дефект нейтрален и может взаимодействовать с водородом. В этом случае модель дефекта описывается как $O_3{\equiv}Si–H$. Предполагается, что E' является центром захвата дырок и характеризуется сечением захвата $3 \cdot 10^{-14}$ cm$^2$. При захвате дырки высвобождается протон, который затем может быть захвачен в окисле, образуя донорные поверхностные состояния и положительный заряд в окисле.

Генерация дефектов оборванных связей кремния в (a)-SiO$_2$ может зависеть от наличия примеси водорода [30]. Водород в (a)-SiO$_2$ разрывает напряженные связи Si–O с относительно небольшой энергией активации (0.5-1.3 eB). Согласно представлениям, развитым в работе [31], водород в системе Si/SiO$_2$ может двигаться продольно в плоскости поверхности кремния, пока не произойдет его захват на субоксидной связи или он не вступит в реакцию со связью Si–H, депассивируя оборванные связи кремния. Диффузией водорода было объяснено одно из первых экспериментальных наблюдений изменения заряда в оксиде кремния под действием ультрафиолетового излучения [32], предполагая взаимодействие водорода с дефектами, присутствующими в окисле до облучения. К дефектам, связанным с водородом, относится мостиковый дефект водорода (–Si–H–Si–), который представляет собой атом водорода, расположенный в центре Si–Si связи [33]. Экспериментальные исследования показали, что дефект мостикового водорода, как и дефект E′, связанный с гидроксильной группой, вносятся технологически. Авторы работы [34] отмечают, что если для формирования –Si–H–Si– дефекта нужна кислородная вакансия в оксиде, то образование гидроксильных E′ дефектов может возникать и в бездефектном аморфном оксиде при высокой температуре, если есть источник водорода. Дефекты окисла существенно влияют на ток утечки, поддерживая подбарьерный перенос носителей заряда между полупроводником и электродом МДП-структуры, а также ухудшают характеристики температурно-полевой стабильности МДП-транзисторов. Экспериментально показано, что водород в оксиде может приводить к формированию (1) термически нестабильных дефектов, которые проявляют себя как донорные состояния и (2) нейтральных электронных ловушек в оксиде при разрыве водородом Si–O связей [35]. Авторы отмечают, что общей причиной, приводящей к генерации дефектов окисла и ГР ДП является инжекция атомарного водорода в оксид и его ионизация на ГР $Si/SiO_2$. Таким образом, механизм деградации изолирующих свойств диэлектрических пленок является электрохимическим.

### 3. Методы анализа параметров МДП-структуры

Поскольку поверхность полупроводника обычно характеризуется достаточно высокой плотностью поверхностных состояний, необходимо рассмотреть методы, использующиеся для определения параметров поверхностных состояний – энергетического распределения плотности поверхностных состояний (ПС) и их сечений захвата.

### 3.1. Метод мультичастотной проводимости

Одним из малосигнальных методов, применяемым в исследовании динамического отклика ОПЗ полупроводника, является метод мультичастотной проводимости. Метод применим как к исследованию объемных центров [36], так и поверхностных состояний [37, 38]. В отношении исследования поверхностных состояний границы раздела полупроводник-диэлектрик, теория метода разработана Николлином и Брюсом [39]. Метод определения параметров поверхностных состояний (ПС) полупроводника на основе анализа мультичастотной проводимости может быть рассмотрен с помощью эквивалентных электрических схем (рис. 6). Рисунок 6(а) показывает эквивалентную схему, описывающую емкость и проводимость МДП-структуры для случая моноэнергетического уровня поверхностных состояний. Схема также учитывает проводимость диэлектрика $G_d$ и последовательное сопротивление $R_s$, которое обусловлено сопротивлением объема полупроводника и контактов. Измеряемая полная проводимость МДП-структуры при некой частоте приложенного напряжения малой амплитуды соответствует некоторому значению полного электрического сопротивления. Из этого значения вычитается реактивная составляющая емкости диэлектрика МДП-структуры, что в представлении МДП-структуры электрической схемой приводит к параллельному соединению емкости обедненного слоя полупроводника и проводимости поверхностных состояний. Рассмотрение непрерывного распределения ПС по временам релаксации и энергии соответствует схеме на рис 6(б), где емкость $C_p$ представляет собой емкость, которая соединена параллельно емкости слоя обеднения полупроводника. Учет влияния последовательного сопротивления может быть проведен на основе эквивалентных схем рис. 6(в, г, д) и описан аналитически:

$$C_m = \frac{C_c}{(G_c R_s + 1)^2 + \omega^2 C_c^2 R_s^2} \qquad (2)$$

$$G_m = \frac{G_c(G_c R_s + 1) + \omega^2 C_c^2 R_s}{(G_c R_s + 1)^2 + \omega^2 C_c^2 R_s^2} \qquad (3)$$

$$C_c = \frac{C_m}{(1 - G_m R_s)^2 + \omega^2 C_c^2 R_s^2} \qquad (4)$$

$$G_c = \frac{\omega^2 C_c C_m R_s - G_m}{(G_m R_s - 1)} \qquad (5)$$

где $C_m$, $G_m$ – значения емкости и проводимости, измеряемые экспериментально, а $C_c$, $G_c$– определенные с поправкой на вклад последовательного сопротивления $R_s$, ω – круговая частота переменного напряжения. Учет последовательного сопротивления особенно важен при изучении поверхностных состояний вблизи разрешенных зон полупроводника, когда частота приложенного напряжения возрастает. Анализ медленных поверхностных состояний вблизи середины запрещенной зоны полупроводника может быть затруднен, если время прохождения основных носителей заряда через диэлектрик и времена релаксации поверхностных центров, принадлежащих полупроводнику, становятся близки [40]. После учета эффектов последовательного сопротивления и проводимости диэлектрика, проводимость поверхностных состояний равна

$$G_{ac} = G_c - G_d \qquad (6).$$

Параметры поверхностных состояний определяются по частотной зависимости проводимости "$G_p/\omega - \log\omega$". При определенной частоте переменного напряжения $\omega\tau = 1$, где $\tau$ – характеристическое время обмена зарядом между ПС и зоной основных носителей заряда, кривая $G_p/\omega - \log\omega$ достигает максимума, который прямо пропорционален плотности поверхностных состояний $D_{it}$. Энергетические распределения сечений захвата $\sigma_{p,n}$ и плотности ПС $D_{it}$ определяются по частотной зависимости проводимости при различных значения поверхностного потенциала $\psi_s$.

Для анализа резонансных кривых проводимости использовались различные представления, учитывающие вероятностный характер распределения заряда окисла и поверхностных состояний [41] и изменение заряда ПС при наличии туннельных токов [42]. Если флуктуации поверхностного потенциала учитываются распределением Гаусса, а плотность поверхностных состояний и сечения захвата постоянны в интервале значений поверхностного потенциала, равного среднеквадратичному отклонению поверхностного изгиба зон, то аналитическое выражение частотной зависимости проводимости поверхностных состояний $G_p$ с учетом флуктуаций поверхностного потенциала $u_s$ принимает вид:

$$\frac{G_p}{\omega} = \frac{qD_{it}(2\pi\sigma_s^2)^{-1/2}}{2\omega\tau} \int\limits_{-\infty}^{\infty} exp[-(u_s - \bar{u}_s)^2/2\sigma_s^2]\ln(1 + \omega^2\tau^2)\, du_s \qquad , (7)$$

где $\bar{u}_s = q\bar{\psi}_s/kT$ – усредненное значение поверхностного потенциала, $\sigma_s$–стандартное отклонение, $\tau$ – время релаксации поверхностных состояний, $D_{it}$ – плотность поверхностных состояний. Экспериментально полученные кривые дисперсии проводимости часто являются ассиметричными, что может быть обусловлено вкладом ПС, характеризующихся существенно отличающимися значениями сечения захвата в узком интервале энергии [43].

Хотя метод мультичастотной проводимости позволяет проводить прямые измерения времен релаксации ПС и плотности ПС, экспериментальные трудности могут возникать, если невозможно точно определить поверхностный потенциал полупроводника, значения инверсионной емкости и емкости диэлектрика, при неполном заполнении уровней ловушек или их высокой концентрации, сравнимой с концентрацией основных носителей заряда. Определение сечений захвата по эмиссионным характеристикам без учета изменения энтропии в эмиссии носителей заряда в разрешенные энергетические зоны [44] и широкое распределение поверхностных состояний по временам релаксации [45] также осложняют интерпретацию экспериментальных результатов. Важным моментом в анализе параметров поверхностных состояний является предположение о том, что времена релаксации поверхностных состояний намного меньше, чем времена релаксации ловушек, распределенных от границы раздела диэлектрик-полупроводник в окисел. Таким образом, предполагаем, что мы исследуем быстрые поверхностные состояния.

Поверхностные состояния границы раздела диэлектрик-полупроводник, принадлежащие окислу, обычно характеризуются большими характеристическими временами. «Быстрые» и «медленные» состояния могут возникать при облучении МДП-структуры фотонами высоких энергий, электронами, а также в сильных электрических полях [46]. Механизмы генерации поверхностных состояний могут быть связаны с электронно-дырочной рекомбинацией в окисле [47], захватом дырочных носителей заряда в диэлектрике («модель водорода» Грискома [48]) или возникновением дефектов оборванных связей в оксиде (например, E´ центров). Механизм генерации ПС будет зависеть от технологической предыстории образца, содержания водорода и механических напряжений в тонких изолирующих пленках на полупроводниках.

Медленные ПС могут изучаться по шумовой составляющей тока МДП-структуры [49]. При анализе отклика ПС методом проводимости и сопоставлении результатов с данными измерений шумов тока, было показано, что быстрые ПС дают вклад в пик потерь проводимости, а медленные – образуют плато при низкой частоте (см. рис. 2 в работе [50]). Состояния в оксиде вносят вклад в 1/f и телеграфный шум [51]. Несколько методик разделения вклада быстрых и медленных состояний ГР ДП были описаны в работах [52, 53, 54]. Работа [54] рассматривает метод определения параметров ПС в широкозонных полупроводниках и использует для этого однопереходный транзистор, чтобы обеспечить достаточную концентрацию неосновных носителей заряда.

**3.2. Емкостные методы анализа параметров МДП-структуры**

Теория квазистатического метода вольт-фарадных характеристик (ВФХ) разработана Берглундом [55]. Квазистатическое приближение предполагает, что частота измеряемого сигнала существенно ниже обратных значений времен релаксации поверхностных состояний и генерационно-рекомбинационных времен основных и неосновных носителей заряда. Метод основан на измерении тока смещения в зависимости от медленно изменяющегося напряжения. Ток смещения пропорционален дифференциальной емкости МДП-структуры при определенном напряжении:

$$i = \frac{dQ}{dt} = \frac{dQ}{dV} \cdot \frac{dV}{dt} = C_{lf}\,\frac{dV}{dt} \qquad (8)$$

где $\frac{dV}{dt}$ – скорость изменения напряжения, $C_{lf}$ – емкость, измеряемая на низкой частоте. Изменение поверхностного потенциала от приложенного напряжения определяется при рассмотрении дифференциального потенциала емкости МДП-структуры:

$$dV = \frac{dV}{C_{ox}} + d\psi_s \qquad (9)$$

и поверхностный потенциал определяется интегрированием выражения

$$\psi_s = \int \left(1 - \frac{C_{lf}}{C_{ox}}\right) dV + \Delta \qquad (10)$$

где $\Delta$ - постоянная интегрирования.

Следует отметить, что метод Термана [56], в котором производится сравнение экспериментальной ВФХ с теоретической, имеет ограниченное применение для определения плотности ПС в случае использования тонких слоев диэлектриков, обладающих высоким значением диэлектрической проницаемости из-за возрастающей неточности в определении потенциала плоских зон [57]. Энергетическое распределение поверхностных состояний по ширине запрещенной зоне полупроводника можно получить, основываясь на изменении заряда поверхностных состояний при изменении положения уровня Ферми с температурой (метод Грея и Брауна) [58], схематически представленного на рис. 7. При понижении температуры уровень Ферми сдвигается к краям разрешенных зон полупроводника, приводя к изменению заряда поверхностных состояний $N_{it}$. Этот заряд $N_{it} = \frac{1}{S}\frac{dQ_{it}}{dE_f}$, где S – площадь электрода, $Q_{it}$ – полный заряд ПС, $E_f$ – уровень Ферми, может быть определен непосредственно по ВФХ, которая сдвигается по напряжению с изменением температуры. Для кремниевых МДП-структур с концентрацией легирующей примеси $10^{14}$-$10^{17}$ см$^{-3}$ поверхностные состояний могут быть исследованы в области энергий 0.02 eB-0.35 eB от краев разрешенных зон. Метод Грея и Брауна так же, как и метод квазистатических ВФХ, не дает возможности разделить поверхностные состояния по временам релаксации и позволяет определить полный заряд быстрых и медленных поверхностных состояний. С использованием МДП-структур на основе кремния электронной и дырочной проводимости, значение полного заряда поверхностных состояний в запрещенной зоне полупроводника $N_{it}$ может быть получено из соотношения для напряжений плоских зон $V_{fb}$, определенных при заданной температуре, например, при температуре жидкого азота: $V_{fbn}$ (77K) – $V_{fbp}$ (77K) = $E_g$ (77K) + $qN_{it}/C_{ox}$, где $q$ – заряд электрона, $C_{ox}$ – емкость диэлектрика в МДП-структуре, $E_g$ – ширина запрещенной зоны полупроводника. Применение данного подхода в определении полного заряда поверхностных состояний имеет то преимущество, что интегрирование в выражении (10) происходит в физическом эксперименте измерением ВФХ.

В качестве примера энергетические распределения поверхностных состояний $D_{it}$, определенное методами Берглунда и мультичастотной проводимости для ГР (100)Si/HfO$_2$, приведены на рис. 8(а). Особенностью спектра является наличие двух максимумов при значениях энергии 0.25 и 0.85 eV выше края валентной зоны кремния, что соответствует переходам (+/0) и (0/-) амфотерного $P_{b0}$ дефекта. Непрерывный спектр поверхностных состояний, существующих из-за наличия слабых связей Si–Si и Si–O на ГР Si/SiO$_2$, имеет U-форму. Авторами работы [59] было сделано предположение, что непрерывное энергетическое распределение U-типа возникает из-за наличия слабых связей в области межфазной границы раздела Si-SiO$_2$. Сечения захвата основных носителей заряда, определенные методом мультичастотной проводимости, показаны на рис. 8(б). В интервале энергий 0.3-0.56 eB значения сечений захвата основных носителей заряда для (100)Si/HfO$_2$ более, чем на порядок, меньше значений, характеризующих $P_{b0}$ дефекты (100)Si/SiO$_2$ интерфейса, что находится в согласии с моделью кулоновского центра, для которого зависимость сечений захвата обратно пропорциональна квадрату диэлектрической постоянной окисла.

Кроме оценки плотности ПС в запрещенной зоне полупроводника, с помощью вольт-фарадных характеристик можно получить распределение концентрации примеси в области обеднения полупроводника. Чтобы минимизировать возможный вклад поверхностных состояний, распределение концентрации примеси определяется по высокочастотной ВФХ, то есть при минимальном влиянии ПС на ВФХ в области обеднения полупроводника основными носителями заряда. Быстрый спад емкости слоя обеднения характерен для низкой концентрации носителей заряда. С ростом концентрации свободных носителей заряда спад ВФХ в области ОПЗ уменьшается [60]. Легирование полупроводника определяется в области обеднения согласно определению ширины области обеднения $W_D$ полупроводника [61]:

$$W_D = \sqrt{\frac{4\varepsilon_s \, kT ln(N_{A,D}/n_i)}{q \cdot (N_{A,D})}}$$

, (11)

где $N_{A,D}$ – концентрация легирующей примеси в полупроводнике, $n_i$ – собственная концентрация носителей заряда при температуре $T$, $q$ – элементарный заряд, $\varepsilon_s$ – диэлектрическая проницаемость полупроводника. Концентрация ионизованной примеси может быть также определена, если известен изгиб зон в полупроводнике при формировании инверсионного слоя $\psi_{s\,inv} \approx \frac{2kT}{q} \ln\left(\frac{N_{A,D}}{n_i}\right)$. Метод определения легирования полупроводника по инверсионной емкости может иметь преимущество, поскольку влияние поверхностных состояний исключено. Данный способ применялся для обнаружения радиолитического водорода в SiO$_2$ при анализе влияния ультрафиолетового облучения на захват заряда в МДП-структурах, имплантированных ионами фтора или аргона [62, 63].

При использовании электрохимического контакта к полупроводнику можно определять распределение концентрации легирующей примеси по толщине пленки в ходе травления. Преимуществом метода является возможность исследования сильно легированных полупроводников [64] и материалов, к которым не могут быть применены измерения на основе эффекта Холла [65]. Пространственное ограничения метода зависит от длины экранирования Дебая в полупроводнике. Перенос заряда в области контакта полупроводника и электролита определяется электрохимическим процессом, и параметры границы раздела полупроводник-электролит (ГР ПЭ) зависят от электронной структуры ГР ПЭ. В отсутствие поверхностных состояний на поверхности полупроводника приложенное напряжение падает на области пространственного заряда полупроводника, что делает возможным определение концентрации ионизированной примеси. Кроме влияния поверхностных состояний, диполи на поверхности полупроводника также могут изменить потенциал плоских зон полупроводникового электрода. В отличие от поверхностных состояний, диполи также изменят и сродство к электрону.

С уменьшением физической толщины окисла до одного нанометра анализ электрических характеристик приборов требует учитывать влияние квантово-механических эффектов области обогащения [66, 67] и инверсии [68] поверхности полупроводника основными носителями заряда. Следует учитывать вклад конечной плотности состояний и конечной толщины инверсионного слоя в инверсионную емкость.

### 3.3. Метод спектроскопии глубоких уровней для исследования параметров поверхностных состояний полупроводника

К нестационарным емкостным методам следует отнести: метод накачки заряда [69], метод нестационарной емкости при импульсном изменении приложенного напряжения [70] и метод спектроскопии глубоких уровней (РСГУ) [71, 72]. Метод релаксационной спектроскопии глубоких уровней был предложен Лангом для исследования центров захвата носителей заряда в области пространственного заряда p-n-переходов и барьеров Шоттки. Изучение релаксации емкости, в отличие от релаксации тока, позволяет исследовать эмиссию как основных, так и неосновных носителей заряда с ловушечных уровней [73]. Техника основана на регистрации сигнала емкости во времени и обработке сигнала интегрированием с определенным временем задержки в течение выбранного периода времени и последующим усреднением по числу измерений. Если энергетический спектр распределения ловушек непрерывен, то измерения дают спектр эмиссионных констант, которые зависят как от распределения ловушек по энергии, так и от сечений захвата. Изначально в методе РСГУ использовались импульсы большой амплитуды, заполняющие поверхностные состояния основными носителями заряда, и емкость переходного процесса, соответствующая релаксации заряда ПС к равновесному значению, регистрировалась в измерении. Для определения энергии ловушки независимо от скорости эмиссии был разработан метод двойных импульсов, которые имеют разную амплитуду. Это позволило заполнять ловушки, распложенные в узком интервале энергий [74]. Еще одна усовершенствованная методика РСГУ была предложена для независимого определения сечений захвата от температуры [75]. Выявление температурной зависимости сечений захвата независимым методом важно, поскольку сильная температурная зависимость сечений захвата характеризует процесс мультифононной эмиссии носителей заряда с поверхностных уровней, как наблюдалось ранее для объемных центров захвата в III-V полупроводниках [76]. Метод основан на использовании импульсов малой амплитуды, чтобы уменьшить энергетический интервал, соответствующий эмиссии носителей заряда с поверхностных состояний. Схематически метод объяснен на рис. 9. Измерения проводятся в режиме постоянной емкости, то есть, ширина слоя обеднения не изменяется, а приложенное напряжение динамически компенсируется во времени $\Delta t = t_2 - t_1$. При приложении последовательности заполняющих импульсов амплитуды $\Delta V$, наложенных на постоянное напряжение, которое переводит поверхность полупроводника в состояние обеднения основными носителями заряда, изменение емкости МДП-структуры за временной интервал $t_2 - t_1$ запишется как

$$\Delta C = A \int_{E_v}^{E_c} N_s(E) \left[ e^{-\frac{t_1}{\tau_n}} - e^{-\frac{t_2}{\tau_n}} \right] [f_0(E) - f_1(E)] dE, \quad (12)$$

где $N_s(E)$ – плотность ПС при энергии $E$, $\tau_n$ – эмиссионное время электронов с ПС, $f_0(E), f_1(E)$ – функции заполнения поверхностных состояний до и после приложения импульса напряжения, заполняющего поверхностные состояния основными носителями заряда. Константа $A$ определяется как $A = C_0^3 / \varepsilon_s C_{ox} N_D$, где $C_0$ – емкость при напряжении, переводящем поверхность полупроводника в состояние обеднения основными носителями заряда, $\varepsilon_s$ – диэлектрическая постоянная полупроводника, $C_{ox}$ – емкость диэлектрика, $N_D$ – концентрация доноров в полупроводнике. Пределы интегрирования ограничены энергиями краев валентной зоны $E_v$ и зоны проводимости $E_c$, $E_f = qV_f$ – энергия Ферми по отношению к краю зоны проводимости $E_c$. Если амплитуда заполняющего импульса мала, функцию заполнения ПС можно считать δ-функцией, и выражение (12) может быть записано как емкость дискретного уровня,

$$\Delta C = A N_s(E_t) \left[ e^{\left(-\frac{t_1}{\tau_n(E_t)}\right)} - e^{\left(-\frac{t_2}{\tau_n(E_t)}\right)} \right] \quad (13)$$

который имеет максимум при $\tau_n = \frac{t_2 - t_1}{ln(t_2/t_1)}$, $E_t$ – энергия предполагаемого дискретного уровня ПС. Время эмиссии выражается через эффективную плотность состояний в зоне проводимости $N_c = N_D e^{(qV_f/kT)}$, тепловую скорость электрона $v_{th}$ и сечение захвата для электрона $\sigma_n$, и эмиссионная постоянная выражается как

$$\tau_n = \left[v_{th} \cdot N_c \cdot \sigma_n e^{(-\Delta E_t/kT)}\right]^{-1}, \qquad (14)$$

$\Delta E_t$ есть энергия активации. Предполагая экспоненциальную зависимость сечений захвата от энергии

$$\sigma_n = \sigma_0 e^{-\Delta E_\sigma/kT}, \qquad (15)$$

где $\sigma_0$ и $\Delta E_\sigma$ – предэкспоненциальный фактор и энергия активации, зависимость сечений захвата от температуры и энергии запишется в виде

$$\sigma_n(E_t, T) = \sigma_0(E_t) e^{(-\Delta E_\sigma(E_t)/kT)}, \qquad (16)$$

и время эмиссии с поверхностных уровней примет вид

$$\tau_n = \left[v_{th} \cdot N_D \cdot \sigma_0(E_t) \cdot exp\left\{-\left[\Delta E_0(E_t) + qV_s - \frac{\Delta E_f}{2}\right]/kT\right\}\right]^{-1}. \qquad (17)$$

Понятно, что выражение (17) в координатах Аррениуса "$ln\,\tau_n - \frac{1}{T}$" позволяет определить энергию активации ловушки по наклону зависимости Аррениуса, а по пересечению с осью ординат – $\sigma_0(E_t)$. Проводя РСГУ измерения при различных значениях поверхностного потенциала, получим значения $\sigma_0(E_t)$ как функцию энергии. Поверхностный потенциал, легирование полупроводника и емкость окисла определяются по ВФХ. Из РСГУ измерений, зная заряд в окисле, плотность и сечение поверхностных состояний, можно определить скорость поверхностной рекомбинации [77]. Изменив полярность приложенных импульсов напряжения, то есть, смещая поверхность полупроводника из состояния обогащения основными носителями в состояние инверсии, определяют скорость термической генерации объемных и поверхностных центров [78], что иногда бывает необходимо при исследовании чистоты технологических процессов. РСГУ метод может быть применен для определения профиля распределения дефектов в объеме полупроводника и диэлектрика. К нестационарным методам также относят методы анализа нестационарной фотоемкости, которые рассматривают кинетику емкости в зависимости от освещенности и смещения на полевом электроде, и оптической релаксационной емкостной спектроскопии.

**3.4. Метод фотоинжекции**

Спектр электронных состояний ГР ДП и диэлектрика может быть исследован с помощью фотоионизации или фотонейтрализации. При фотовозбуждении освобождение носителей с уровней ловушек заряда вносит вклад в фототок. Однако фототок пропорционален плотности состояний и сечению фотоионизации, зависящему от энергии, что делает измерение фототока неприменимым для получения информации о плотности электронных состояний. В связи с этим, в работе [79] был развит подход, использующий измерение изменения заряда в диэлектрике при оптическом возбуждении. Определение спектра электронных состояний основано на оптическом возбуждении электронных переходов из заселенных уровней ловушек заряда в зону проводимости диэлектрика. Изменение заряда в диэлектрике определяется по высокочастотной ВФХ. Время, выбирающееся для высвобождения носителей заряда с ловушечных уровней, должно быть достаточным для практически полного опустошения ловушек вблизи порога ионизации, что дает возможность минимизировать ошибку в энергетическом распределении электронных состояний. В эксперименте по фотоинжекции заряд, вводимый в оксид (интеграл тока по времени инжекции), должен оставаться постоянным, то есть не изменяться методом, который применяется для измерения заряда. Электрическое поле, создаваемое зарядом захваченных носителей, изменяет приповерхностный изгиб зон полупроводника, то есть, объемный заряд полупроводника служит чувствительным элементом наведенного электрического поля. Приповерхностный изгиб зон полупроводника в зависимости от электрического поля может быть определен из ВФХ, а вклад захваченного заряда – по сдвигу ВФХ по напряжению. В МДП-транзисторах захваченный заряд может контролироваться как функция порогового напряжения. Эта методика определяет плотность носителей заряда в инверсионном канале для контроля электрического поля на ГР ДП. Кроме того, электрическое поле, наведенное захваченным зарядом, может определяться методом Кельвина или фотоэдс. В последнем случае интенсивность света должна быть достаточной для достижения условия плоских зон на поверхности полупроводника. В работе [80] метод фотоопустошения ловушек был применен для исследования электронных центров захвата в $SiO_2$ после имплантации ионами натрия и алюминия. Методы, основанные на фотоинжекции заряда в структурах металл-полупроводник, обладают чувствительностью к локальной неоднородности заряда в области барьера, поскольку заряд в полупроводнике индуцирует равный заряд в электродах, обуславливая появления электрического поля в области границы раздела и соответствующего изменения высоты барьера металл-полупроводник [81, 82]. Для определения пространственного распределения заряда в многослойных диэлектриках (приповерхностный заряд или объемный заряд в диэлектрике) может использоваться изменение толщины приповерхностных подслоев на полупроводниковой подложке травлением под наклоном [83].

**4. Оксид гафния в качестве диэлектрика**

Электронные свойства новых диэлектриков для использования в комплементарной технологии транзисторов изучались особенно интенсивно на протяжении последних десятилетий [84]. Среди разнообразия материалов, обладающих высокой диэлектрической проницаемостью, был выбран оксид гафния $HfO_2$, а в качестве метода формирования тонких пленок на полупроводниках — метод молекулярного наслаивания [85]. Особенности свойств оксида гафния обусловлены электронной структурой гафния — наличием заполненной внутренней f-оболочки, что определяет химические свойства его соединений. Наличие f электронов у элементов третьей переходной группы (от La до Hf) оказывает влияние на электронную структуру, определяемую 5d и 6s орбиталями. В частности, в отличие от $ZrO_2$, $HfO_2$ не образует силицидов, что важно для формирования диэлектрика на кремнии [86]. В настоящее время отмечается интерес к $HfO_2$ и $ZrO_2$ как к ферроэлектрикам, в которых ферроэлектрическое состояние возможно при возникновении спонтанной поляризации, если достигается кристаллизация в орторомбической фазе, а не в моноклинной. Отмечалась стабильность электрических свойств $HfO_2$ по отношению к высокотемпературному отжигу, низкие плотности тока при уменьшении толщины диэлектрика до субнанометровых значений. Показано, что фиксированный заряд в оксиде и захват заряда все еще являются сдерживающим фактором в увеличении подвижности носителей заряда в полевых транзисторах и достижении нужных значений порогового напряжения. Природа захваченного заряда в $HfO_2$ исследовалась в работах [87, 88]. При изучении тонкой структуры спектров кислорода методом спектроскопии характеристических потерь в работе [87] было предположено, что при высокотемпературном отжиге уменьшение точечных дефектов в $HfO_2$ сопровождается увеличением координационного числа гафния. Центроид заряда в $HfO_2$ сосредоточен вблизи границы раздела с полупроводником. В согласии с данным наблюдением находятся результаты по исследованию захвата заряда в кремниевых МДП-структурах с различными диэлектриками ($Al_2O_3$, $ZrO_2$, $HfO_2$). Показано, что положительный захваченный заряд находится в подслое $SiO_x$, который выращен на кремнии с целью обеспечить двумерный рост диэлектрика или сформирован в процессе осаждения диэлектрика. Захват дырок является преобладающим процессом захвата заряда в исследованных диэлектриках и коррелирует с высвобождением протонов в процессе генерации электронно-дырочных пар фотонами энергии 10 eB, предполагая существенный вклад протонов в захваченный в окисле положительный заряд. В экспериментах по захвату положительного заряда в окисле высвобождение водорода регистрировалось по дезактивации примеси бора в кремнии. Изменение работы выхода полевого электрода влияло на величину захваченного положительного заряда в окисле, говоря о том, что дефекты окисла принимают участие в токопереносе через диэлектрик.

Электронный захват в окиси гафния детально рассмотрен в работе [89]. Методом фотоопустошения ловушек обнаружено два сорта центров захвата заряда при энергиях $E_t$ =2.0 eB и $E_t$ =3.0 eB ниже дна зоны проводимости $HfO_2$. Оба сорта ловушек не чувствительны к методу формирования окиси гафния и примеси, внедренной в диэлектрик в процессе осаждения. Кроме того, захват на акцепторные состояния не коррелирует с механизмом захвата заряда, характерным для изолированной вакансии кислорода, предполагая, что данный сорт акцепторных центров – собственные дефекты матрицы оксида. Поляронный характер токопереноса носителей заряда известен как для аморфных слоев оксида гафния [90], так и для оксида кремния [48]. Если в аморфном $SiO_2$, полученным термическим окислением кремния, поляроны стабильны при низких температурах, то в слоях окислов металлов поляронные эффекты могут наблюдаться при более высоких температурах, поскольку современные технологии используют низкотемпературные процессы получения тонких пленок на полупроводниках.

Что же касается спектра поверхностных состояний ГР кремний-оксид гафния, то здесь существенно влияние технологии получения диэлектрика. Так, самая низкая плотность ПС характерна для ГР, сформированных молекулярным наслаиванием из тетрахлорида гафния и воды на предварительно окисленном в озонированной воде кремнии. Использование мос-гидридной технологии получения $HfO_2$ показывало более высокие плотности ПС. Нитридизация как в процессе формирования подслоя оксинитрида или нитрида кремния, так и использование азото-содержащего реагента при росте оксида гафния приводило к существенному увеличению ПС и формированию ПС, связанных с внедрением азота в приповерхностный слой окисла на кремнии [91]. Для ГР ДП окислов $Al_2O_3$, $ZrO_2$, $HfO_2$ и (100)Si характерны типичные дефекты оборванных связей кремния – $P_{b0}$ [92], пассивация которых зависит от катиона металла, формирующего интерфейс с кремнием [93].

**5. Выводы**

Приведено аналитическое рассмотрение методов, использующихся при изучении электронных свойств тонких пленок на кремнии в МДП-структуре. Показано, что часть точечных дефектов системы $Si-SiO_2$ также наблюдаются на ГР, сформированных тонкими пленками окислов переходных металлов на кремнии. В последнем случае дефекты оборванных связей в системе кремний-изолятор не являются единственным источником заряда поверхностных состояний и захваченного заряда в окисле, предполагая в последнем случае влияние поляронного механизма захвата заряда в тонких пленках окислов металлов. Описанные в работе методики являются применимыми к исследованию многих других систем диэлектрик-полупроводник на основе неорганических и органических материалов электронной техники, хотя и получили развитие в ходе совершенствования кремниевой технологии МДП-транзисторов.

[1] Rieger P H *Electron Spin Resonance: Analysis and Interpretation* (Cambridge: RSC Publishing, 2009)

[2] Stesmans A *J Magnetic Resonance* **76** 14 (1988)

[3] Nishi Y *Jpn. J. Appl. Phys.* **10** 52 (1971)

[4] Stesmans A *Appl. Phys. Lett.* **48** 972 (1986)

[5] Stesmans A , Afanaas'ev V V *J. Appl. Phys.* **83** 2449 (1998)

[6] Caplan P J et al. *J. Appl. Phys.* **50** 5847 (1979)

[7] Poindexter E H et al. *J. Appl. Phys.* **52** 879 (1981)

[8] Poindexter E H et al. *J. Appl. Phys.* **56** 2844 (1984); Gerardi G J et al. *Appl. Phys. Lett.* 1986. **49**. 348 (1984); Chang S T, Wu J K and Lyon S A *Appl. Phys. Lett.* **48** 662 (1986)

[9] Johnson N M, Shan W, and Yu P Y, *Phys. Rev. B* **39** 3431(R) (1989)

[10] Grimmeiss H G *Phys. Rev. B*. **39** 5175 (1989)

[11] Stesmans A and Afanas'ev V V *Phys. Rev. B* **57** 10030 (1998)

[12] Nishi Y K and Ohwada A *Jpn. J. Appl. Phys.* **11** 85 (1972)

[13] Kolomiiets N M et al. *ECS J. Solid State Sci. Technol.* **5** Q3008 (2016)

[14] Xu N et al. *Appl. Phys. Lett.* **107** 123502 (2015)

[15] Brower K L *Phys. Rev. B* **33** 4471 (1986)

[16] Stesmans A and Braet J in *Insulating Fims on Semiconductors*, ed. Simone J J and Buxo J (North Holland, Amsterdam, 1986) c. 25

[17] Pierreux D and Stesmans A *Phys. Rev. B* **66** 65320 (2002)

[18] Stesmans A *Phys. Rev. B* **48** 2418 (1993)

[19] Lucovsky G et al. *J. Vac. Sci. Technol. B* **17** 1806 (1999)

[20] Stesmans A and Scheerlinck F Phys. Rev. B 50 5204 (1994); Stesmans A, Scheerlinck F, and Afanas'ev V, *Appl. Phys. Lett*. **64** 2282 (1994)

[21] Stapelbroek M et al. *J. Non-Cryst. Solids* **32** 313 (1979)

[22] Friebele E J *Phys. Rev. Lett.* **42** 1346 (1979)

[23] Edwards A H and Fowler W B, *Phys. Rev. B* **26** 6649 (1982)

[24] Uchino T, Takahashi M, and Yoko T *Phys. Rev. Lett.* **86** 4560 (2001)

[25] Griscom D L *Phys. Rev. B* **20** 1823 (1979)

[26] Uchino T, Takahashi M, and Yoko T *Phys. Rev. Lett.* **86** 5522 (2001)

[27] Griscom D L and Friebele E J *Phys. Rev. B* **34** 7524 (1986)

[28] Vanheusden K and Stesmans A *J. Appl. Phys*. **74** 275 (1993)

[29] Afanas'ev V V and Stesmans A *J. Phys.: Condens. Matter*. **12** 122285 (2000)

[30] El-Sayed Al-M et al. *Phys. Rev. Lett*. **114** 115503 (2015)

[31] Rashkeev S N et al. Phys. Rev. Lett. **87** 165506 (2001)

[32] Weinberg Z A, Rubloff G W, Bassous E *Phys. Rev. B*. **19** 3107 (1979)

[33] P.E. Blöchl, Phys. Rev. B 62 6158 (2000)

[34] Grasser T et al. *Electron Devices Meeting*, IEDM14-530 (2014)

[35] Afanas'ev V V and Stesmans A *J Electrochem Soc* **146** (1999)

[36] Cordaro J F, Shim Y, and May J E *J. Appl. Phys*. **60** 4186 (1986)

[37] Uren M J, Brunson K M, and Hodge A M *Appl. Phys. Lett*. **60** 624 (1992)

[38] Pensl G et al. *Microelectronic Engineering*. **83** 146 (2006)

[39] Nicollian E H and Brews J R (Wiley, 2002)

[40] Vogel T V, Henson W K, Richter C A *IEEE Transactions on Electron Devices* **47** 601 (2000)

[41] Nicollian E H, Goetzberger A *Bell Labs Technical Journal* **XLVI** 1055 (1967)

[42] Preier H *Appl. Phys. Lett*. **10** 361 (1967)

[43] Collins S, Kirton M J, and M. J. Uren M *J Appl. Phys. Lett*. **57** 372 (1990)

[44] Kirton M J and Uren M J *Appl. Phys. Lett*. **48** 1270 (1986)

[45] Penumatcha A V, Swandono S, and Cooper J A *IEEE Transactions on ED* **60** 923 (2013)

[46] Fleetwood D M *Microelectronics Reliability* **42** 523 (2002)

[47] Lai S K *J. Appl. Phys* **54** 2540 (1983)

[48] Griscom D L J. Non-Cryst. Sol. **352** 2601 (2006)

[49] Christensson S, Lundström I, Svensson C *Solid-State Electronics* **11** 797 (1968)

[50] Uren M J, Coilins S, and Kirton M J *Appl. Phys. Lett*. **54** 1448 (1989)

[51] Kirton M J, Uren M J *Advances in Physics*. **38:4** 367 (1989)

[52] McWhorter P J and Winokur P S *Appl. Phys. Lett*. **48** 133 (1986)

[53] Paulsen R E et al. *IEEE Electron Device Letters* **13** 627 (1986)

[54] Sheppard S T, Melloch M R, and Cooper J A *IEEE Transactions on Electron Devices* **41** 1257 (1994)

[55] Berglund C N *IEEE Transactions on Electron Devices* **3** 701 (1966)

[56] Terman L M *Solid-State Electron*. **5** 285 (1962)

[57] Vogel E M, Sonnet A M, Hinkle C L *ECS Trans*. 11(4) 393 (2007)

[58] Gray P V, Brown D M, *Appl. Phys. Lett*. **8** 31 (1966)

[59] Sakurai T and Sugano T *J. Appl. Phys*.1981;**52**:2889. DOI: 10.1063/1.329023

[60] Kennedy D P, Murley P C, Kleinfelder W *IBM Journal of Research and Development* **12** 399 (1968)

[61] Sze M *Physics of Semiconductor Devices* (Wiley, New York, 1969) c. 432-436

[62] Afanas'ev V V et al. *J. Appl. Phys*. **78** 6481 (1995)

[63] Afanas'ev V V, de Nijs J M M, and Balk P *J. Appl. Phys*. **76** 7990 (1994)

[64] Blood P *Semicond. Sci. Technol*. **1** 7 (1968)

[65] Koeder A et al. *Appl. Phys. Lett*. **82** 3287 (2003)

[66] Rios R and Arora N D *IEDM '94. Technical Digest* (1994)

[67] Rana F, Tiwari S, and Buchanan D A *Appl. Phys. Lett*. **69** 1104 (1996)

[68] Takagi S, Toriumi A *IEEE Transactions on Electron Devices* **42** 2125 (1995)

[69] Brugler J S, Jespers P G A *IEEE Transactions on ED* **16** 297 (1969)

[70] Heiman F P IEEE Transactions on ED **14** 781 (1967)

[71] D. K. Schroder *Semiconductor Material and Device Characterization* (3rd ed. Wiley 2006)

[72] Yamasaki K, Yoshida M, Sugano T, *Jap. J. Appl. Phys.* **18** 113 (1979)

[73] Dobaczewski L, Peaker A R, and Bonde Nielsen K *J. Appl. Phys*. **96** 4689 (2004)

[74] Johnson N M *Appl. Phys. Lett*. **34** 802 (1979)

[75] Katsube T, Kakimoto K, and Ikoma T *J. Appl. Phys*. **52** 3504 (1981)

[76] Schulz M, Johnson N M *Solid State Comm.* **25** 481 (1978)

[77] Aberle A G, Glunz S and Warta W *J. Appl. Phys*. **71** 4422 (1992)

[78] Khorasani A E, Schroder D K, Alford T L **61** 3282 (2014)

[79] Afanas'ev V V and Stesmans A *Phys. Rev. B* **59** 2025 (1999)

[80] Harari E, Royce B S H *Appl. Phys. Lett*. **20** 288 (1973)

[81] DiStefano T H *Appl. Phys. Lett*. **19** 280 (1971)

[82] Afanas'ev V V *Internal Photoemission Spectroscopy. Principles and Applications* (Elsevier ed. 2008)

[83] Kaushik V S et al. IEEE Transactions on ED **53** 2627 (2006)

[84] He G, Sun Zh *High-k Gate Dielectrics for CMOS Technology* (Wiley 2012)

[85] Leskela M, Ritala M *Thin Solid Films* **409** 138 (2002)

[86] Zheng W et al. *J. Phys. Chem. A.***109** 11521 (2005)

[87] Wilk G D and Muller D A *Appl. Phys. Lett*. **83** 3984 (2003)

[88] Afanas'ev, V V and Stesmans A *J. Appl. Phys*. **95** 2518 (2004)

[89] Cerbu F et al. *Appl. Phys. Lett*. **108** 222901 (2016)

[90] Kaviani M et al. Phys. Rev. B. **94** 020103(R) (2016)

[91] Fedorenko Y G  J. Appl. Phys. **98** 123703 (2005)

[92] Stesmans A. and Afanas'ev V V *Appl. Phys. Lett*. **80** 1957 (2002)

[93] Truong L *Microelectronics Reliability* **45** 823 (2005)

Рисунок 1.

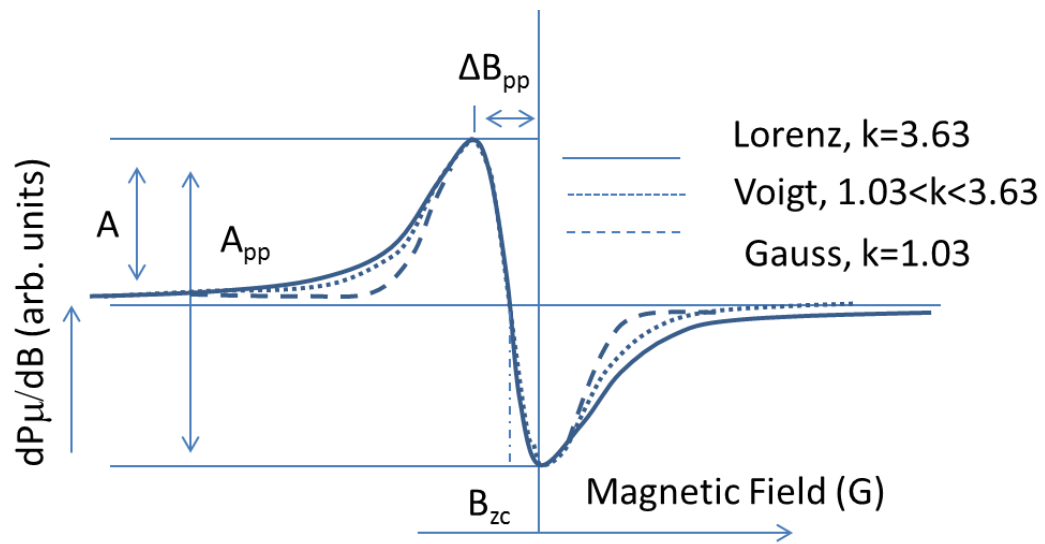

Рисунок 2.

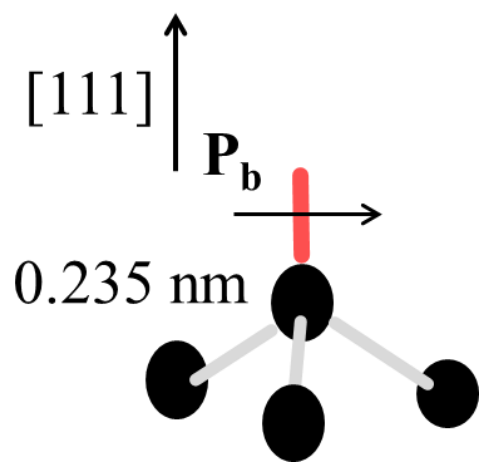

Рисунок 3.

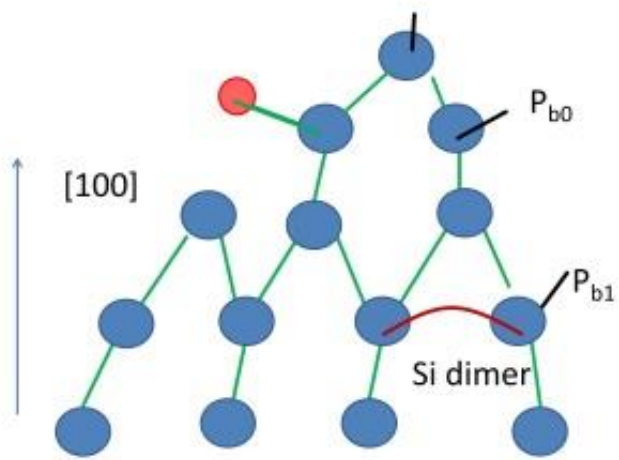

Рисунок 4.

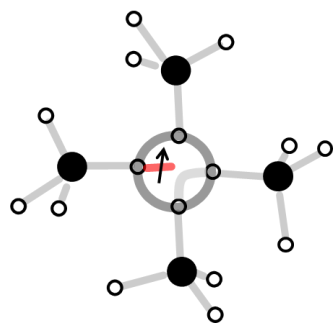 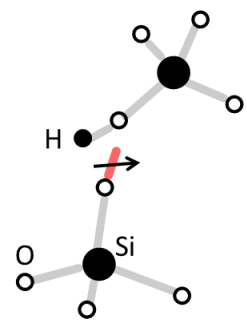

(а) (б)

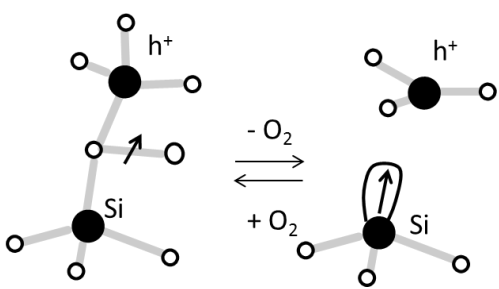 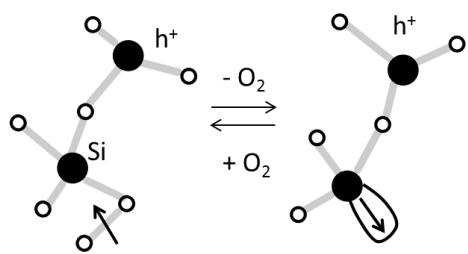

(в) (г)

Рисунок 5.

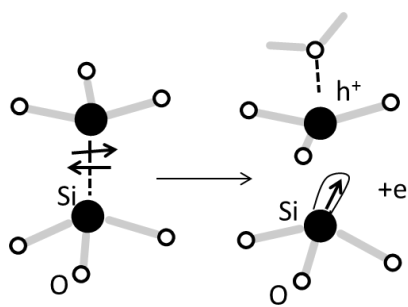 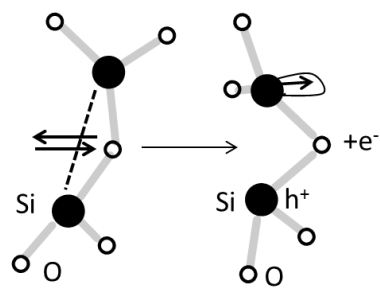

(а)                                                        (б)

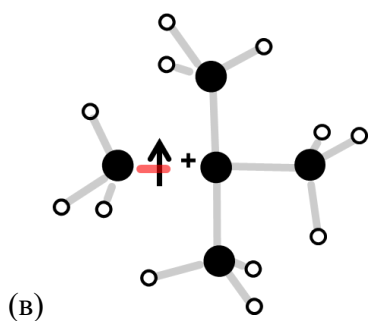

(в)

Рисунок 6.

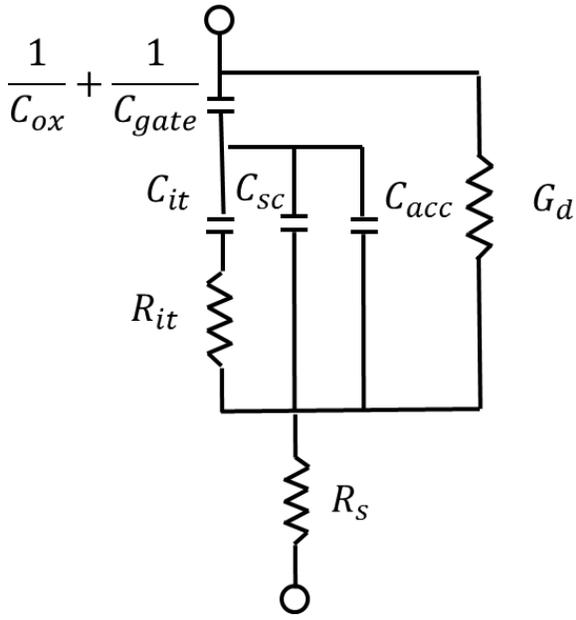
(а)

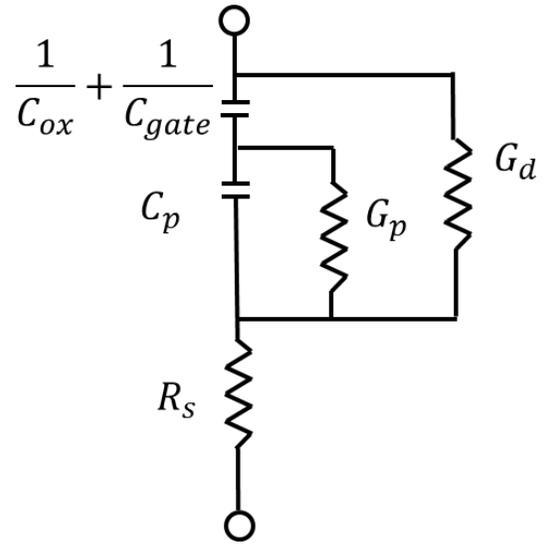
(б)

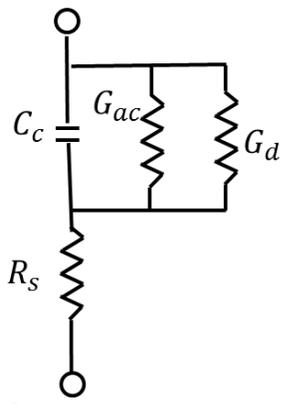
(в)

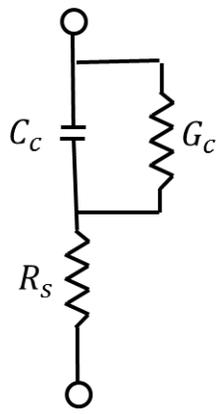
(г)

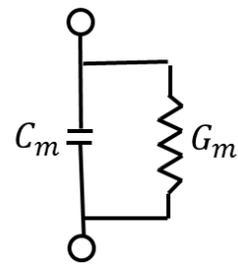
(д)

Рисунок 7.

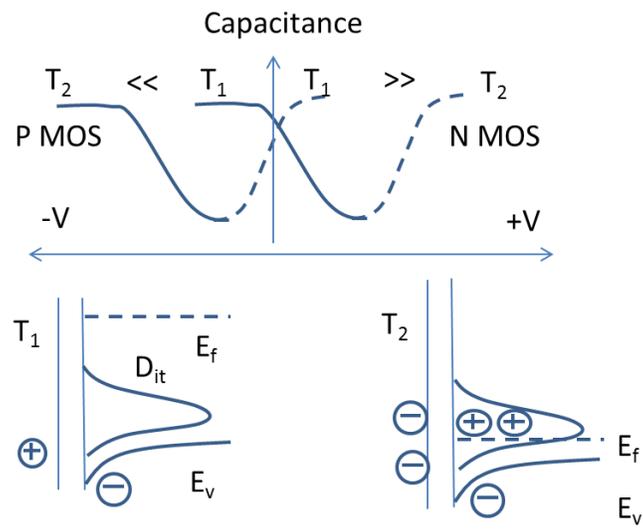

Рисунок 8.

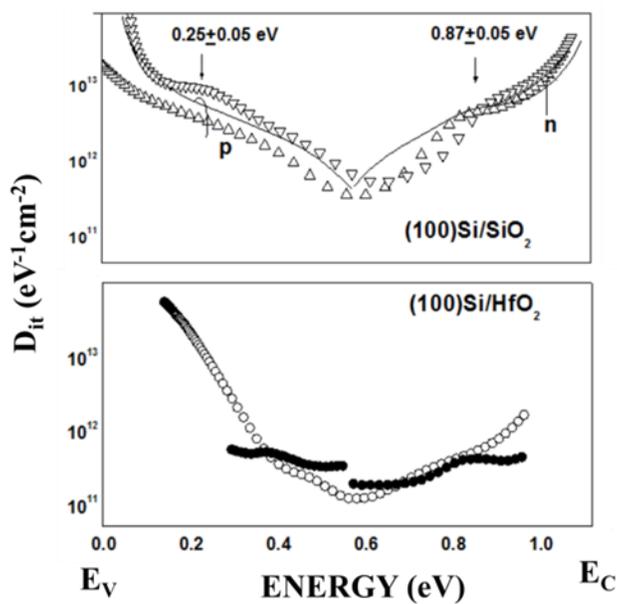

(а)

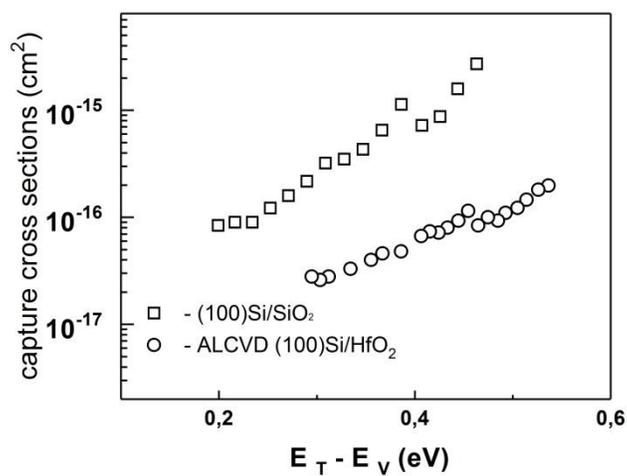

(б)

Рисунок 9.

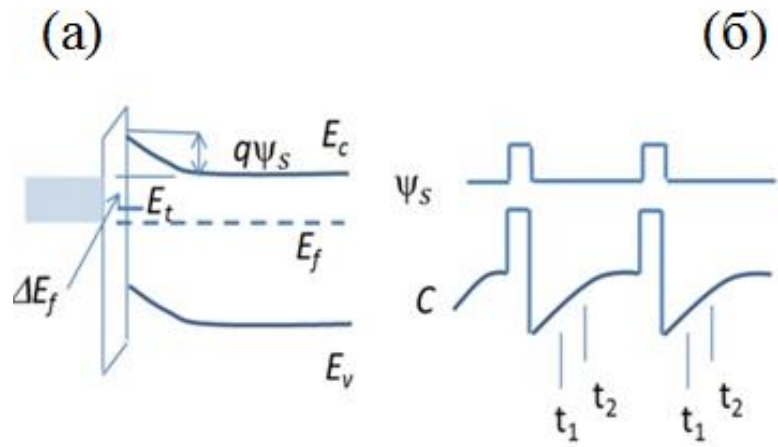

Подписи к рисункам

Рисунок 1. Первая производная dPµ/dB поглощения микроволнового излучения. $B_{zc}$ – значение магнитного поля, при котором dPµ/dB = 0. $\Delta B_{pp}$ – удвоенная амплитуда сигнала.

Рисунок 2. Схематическое изображение $P_b$ центра границы раздела (111)Si/SiO$_2$.

Рисунок 3. Конфигурация $P_{b0}$ и $P_{b1}$ дефектов ГР (100)Si/SiO$_2$.

Рисунок 4. Модели EX центра (а), немостикового кислорода (б) и пероксид-радикала Si–O–O• в предположении существования напряженной (в) и ненапряженной (г) Si–Si связи.

Рисунок 5. Первая модель E´$_\gamma$ центра (а), модель мостикового бескислородного центра захвата дырок (б), модель E´$_\delta$ дефекта согласно работе [28] (в).

Рисунок 6. Эквивалентная схема МДП-структуры для объяснения метода мультичастотной проводимости. (а) общая схема с учетом емкости МДП-структуры в состоянии обеднения и обогащения полупроводника основными носителями заряда для случая моноэнергетического уровня поверхностных состояний: $\frac{1}{C_{ox}} + \frac{1}{C_{gate}}$ – емкость окисла и электрода, $G_d$ – проводимость МДП-структуры с учетом механизма проводимости диэлектрика с участием поверхностных состояний и процессов генерации-рекомбинации в полупроводнике, $C_{it}$, $R_{it}$ – емкость и сопротивление поверхностных состояний, $C_{sc}$ – емкость обедненного слоя полупроводника, $C_{acc}$ – емкость слоя обогащения полупроводника основными носителями заряда, $R_s$ – последовательное сопротивление, учитывающее вклад объема полупроводника и контактов. (б) схема (а) для случая непрерывного распределения ПС по временам релаксации и энергии. (в) схема (б), показывающая емкость $C_c$ с поправкой на последовательное сопротивление, и проводимость $G_{ac}$. (г) схема (в), показывающая емкость $C_c$ и $G_c$ с поправкой на последовательное сопротивление. (д) схема (г), показывающая измеряемые емкость $C_m$ и проводимость $G_m$. (Согласно работе [40]).

Рисунок 7. Схематическое представление метода Грея и Брауна, иллюстрирующее сдвиг уровня Ферми при изменении температуры.

Рисунок 8. (а) Энергетическое распределение поверхностных состояний, определенное методом Берглунда (∇,Δ,–,○) и методом мультичастотной проводимости (•), для

(100)Si/SiO$_2$ и (100)Si/HfO$_2$ и (б) сечений захвата, определенных для тех же образцов (согласно работе [91])

Рисунок 9. Схематическая диаграмма энергетических зон на поверхности полупроводника (а), кинетики емкости $C$ и поверхностного потенциала $\psi_s$ (б) для пояснения метода РСГУ в режиме постоянной емкости при использовании малых импульсов, заполняющих ПС основными носителями заряда.